\documentclass{appolb}
\usepackage{graphicx}
\usepackage{amsmath}
\usepackage{amssymb}
\usepackage{bm}
\usepackage{color}
\usepackage[sort&compress]{natbib}
\setcitestyle{numbers,open={[},close={]},comma}
\usepackage{hyperref}
\def\qhat{\hat{q}}
\begin{document}
\title{Transport properties of the QCD medium%
\thanks{Presented at Quark Matter 2022}%
}
\author{Jacopo Ghiglieri
\address{SUBATECH, Universit\'e de Nantes, IMT Atlantique, IN2P3/CNRS,\\
4 rue Alfred Kastler,
La Chantrerie BP 20722, 44307 Nantes, France}
}
\maketitle
\begin{abstract}
    I present an overview of recent developments 
    in the microscopic description of 
    the quark-gluon plasma. I will concentrate on \emph{medium-induced emission}  and 
    \emph{transverse momentum broadening}. These are two key ingredients of 
    the theory of jet modifications in the QCD medium 
    and of the  kinetic theory used for transport 
    and thermalisation.
  The main focus is on progress towards a better understanding of theory and of its 
    uncertainties. 
\end{abstract}

As per the abstract, my main driver in discussing recent developments 
will be connecting the 
microscopic description to a better understanding and control of theory
uncertainties. 
 For a connection of these developments to jet quenching
data --- summarised at this conference in~\cite{leticia,rybar} ---
I refer to~\cite{jasmine}. Thermalisation and hydrodinamisation have 
been discussed in~\cite{spalinski}.
\section{Medium-induced emission (MIE)}
Hard jet partons of energy $E$ much larger 
tha the temperature $T$ propagating through the QCD medium experience
frequent soft interactions with the medium, which 
can eventually source collinear, medium-induced radiation. 
Its long formation time  make it sensitive to the 
Landau--Pomeranchuk--Migdal (LPM) effect, i.e.  the
quantum-mechanical interference of many soft scatterings.
 These effective $1\leftrightarrow 2$
processes are not only key in jet modification; through their 
number-changing nature  they
ensure chemical equilibration and energy transport in \emph{bottom-up
thermalisation} \cite{Baier:2000sb}.

The celebrated BDMPS-Z \cite{Baier:1996kr,Zakharov:1996fv}
MIE probability, i.e.
\begin{equation}
    \frac{dI}{dx}=\frac{{\alpha_s P_{1\to 2}(x)}}{[x(1-x)E]^2}
    \mathrm{Re}\int_{t_1<t_2}dt_1dt_2 {{\bm\nabla}_{\boldsymbol{b}_2}\cdot 
    {\bm\nabla}_{\boldsymbol{b}_1}}\bigg[
        \big\langle {\boldsymbol{b}_2},t_2\vert {\boldsymbol{b}_1},t_1
        \big\rangle_{ \boldsymbol{b}_2=0}^{ \boldsymbol{b}_1=0}-
        \text{vac}\bigg],
\end{equation}
is factorised in a DGLAP splitting kernel $P_{1\to 2}(x)$
multiplying a propagator $ \big\langle {\boldsymbol{b}_2},t_2\vert {\boldsymbol{b}_1},t_1
\big\rangle$ that describes diffusion in transverse position space
from the emission in the amplitude at time $t_1$ and vanishing $\boldsymbol{b}$ 
 to the emission in the conjugate amplitude at time $t_2$.
This propagator is a Green's function of 
\begin{equation}
    \mathcal{H}={-\frac{\nabla^2_{\boldsymbol{b}}}{2x(1-x)E}
    +\sum_{i}\frac{{m_i^2}}{2E_i}}
    {-i\mathcal{C}(\boldsymbol{b},x \boldsymbol{b}, (1-x) \boldsymbol{b})}.
    \label{prop}
\end{equation}
This 2D Hamiltonian has a real kinetic term
with in-medium masses $m_i$. The imaginary part is the 
\emph{scattering kernel}. It encodes jet-medium interactions 
and it is tightly related to transverse momentum broadening (TMB). 

Determining the propagator from Eq.~\eqref{prop} is  difficult. 
Historically, approaches concentrated on a few limiting
cases. For thin media one can truncate the LPM resummation series 
at first order in the \emph{opacity expansion}~\cite{Gyulassy:1999zd},
while for thick media one can perform the 
\emph{harmonic oscillator approximation}, introducing the 
\emph{momentum broadening coefficient} $\hat{q}$, i.e.  
${\mathcal{C}(\boldsymbol{b},x \boldsymbol{b}, (1-x) \boldsymbol{b})}
\approx \mathcal{C}_\mathrm{HO}\equiv \frac{{\hat{q}}}{4}\big[b^2+(xb)^2+((1-x)b)^2\big]$ \cite{Baier:1996kr}.
The determination of the propagator simplifies also in 
an infinite, static medium~\cite{Arnold:2002ja}.

\begin{figure}[t]
    \begin{center}
        \includegraphics[width=7.5cm]{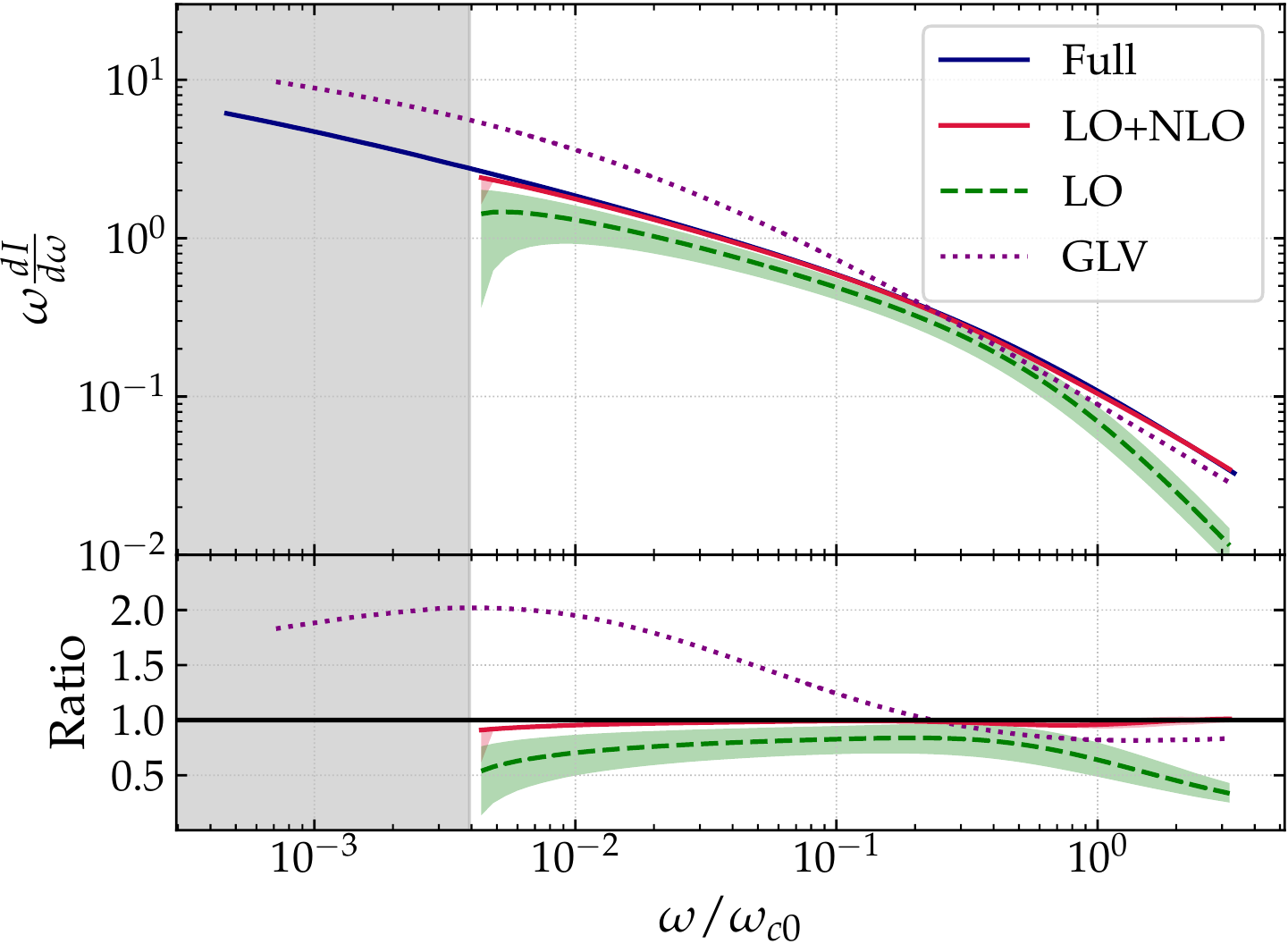}
    \end{center}
    \caption{The ``full'' line is the numerical solution from~\cite{Andres:2020vxs,Andres:2020kfg}.
    The ``LO+NLO'' is the improved opacity expansion, LO is the harmonic oscillator and GLV the 
    first order in opacity.
    Figure taken from~\cite{Barata:2021wuf}.
    \label{fig:ioe}}
\end{figure}
The \emph{improved opacity expansion}, introduced in \cite{Mehtar-Tani:2019ygg,Barata:2021wuf},
is an economical, analytical prescription for overcoming these limiting cases.
It corresponds 
to treating the non-harmonic parts of the scattering kernel as perturbations,
i.e. $\mathcal{C}=\mathcal{C}_\mathrm{HO}+
[\mathcal{C}-\mathcal{C}_\mathrm{HO}]$. In this way, the Coulomb logarithm,
$\mathcal{C}(\boldsymbol{b})\propto b^2\ln(b)$, is captured, thus incorporating
the rarer harder \emph{Molière scatterings}. 
This new prescription was recently tested against 
new numerical determinations of the full propagator, obtained 
in~\cite{Andres:2020vxs,Andres:2020kfg,carlota}, finding good, 10\% or better 
agreement over its validity range, as shown in Fig.~\ref{fig:ioe}.

\section{Transverse momentum broadening}
The discussion so far was agnostic to the specifics of the TMB kernel $\mathcal{C}$
and associated probability $\mathcal{P}(k_\perp)$, other than featuring
a $1/k_\perp^4$ Coulomb tail in the UV and a diffusive gaussian in the IR. 
If we were to determine the TMB kernel perturbatively, we would 
run into the known issues associated with the Linde problem: soft 
gluons ($\omega\ll T$) are \textit{classical} high-occupancy modes, distributed on the $T/\omega$ IR 
tail of the Bose distribution. As the expansion parameter becomes $g^2 T/\omega$, 
convergence can be seriously hampered.

A breakthrough came from the realisation in~\cite{CaronHuot:2008ni} that these soft 
classical modes at space-like separations become Euclidean. 
As $\mathcal{C}$ is determined 
from  Wilson lines at space-like separations, 
the large-distance contribution $b\gtrsim 1/gT$ can be determined 
non-perturbatively using the dimensionally-reduced theory on the lattice,
\cite{Panero:2013pla,Moore:2019lgw}. At shorter distance one can use pQCD 
and merge the two~\cite{Moore:2021jwe,Schlichting:2021idr,soudi}.\footnote{
The same method is also being applied to in-medium masses~\cite{Moore:2020wvy,Ghiglieri:2021bom,philipp}.}
\begin{figure}[t]
    \begin{center}
        \includegraphics[width=6.8cm]{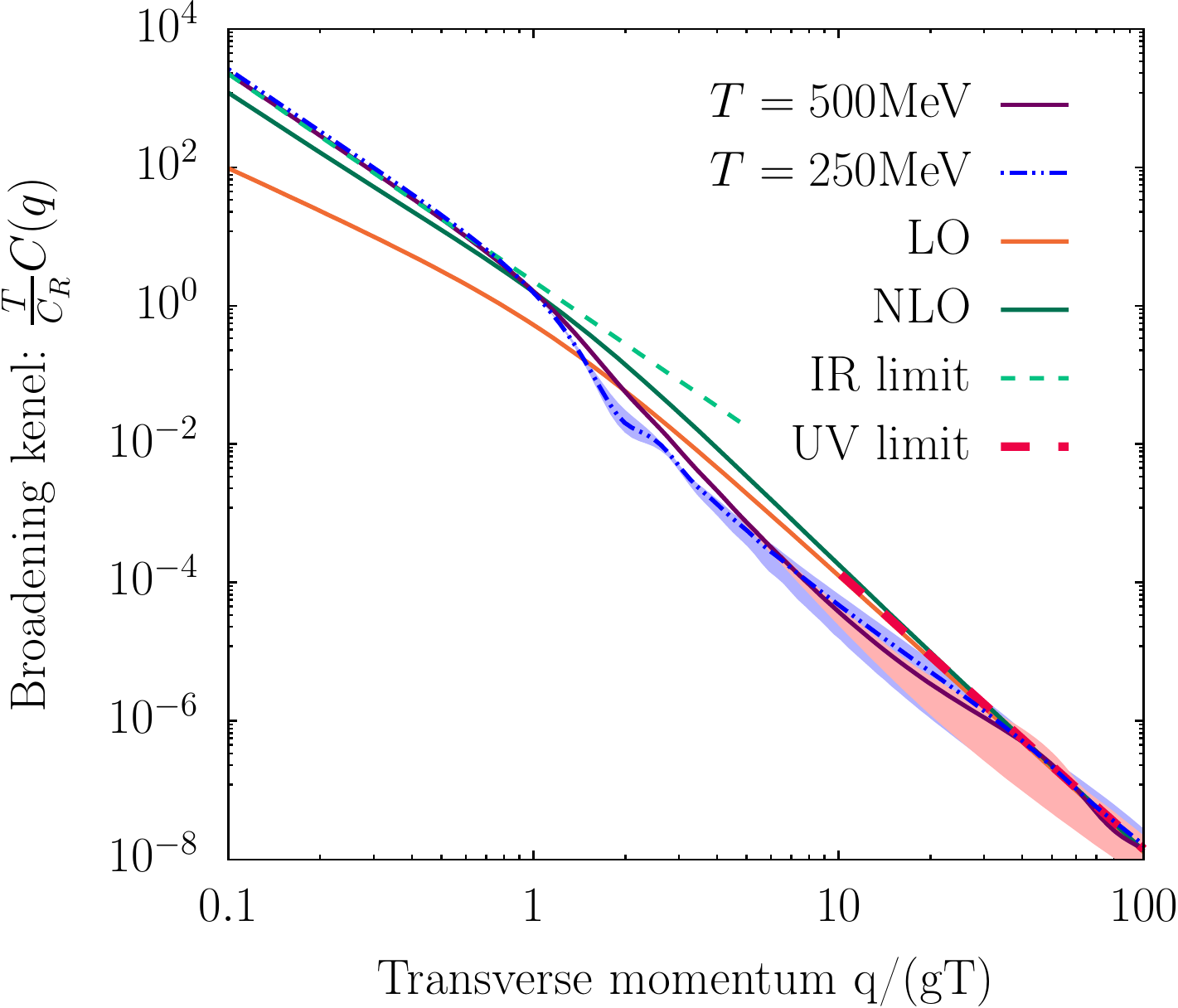}
    \end{center}
    \caption{The LO perturbative line comes from \cite{Aurenche:2002pd,Arnold:2008vd},
    the NLO includes the $\mathcal{O}(g)$ corrections from~\cite{CaronHuot:2008ni}
    following~\cite{Ghiglieri:2018ltw}.
    Figure taken from~\cite{Schlichting:2021idr}.\label{fig:kernel}}
\end{figure}

The resulting kernel is shown in Fig.~\ref{fig:kernel} from~\cite{Schlichting:2021idr}: 
a transition from $1/q_\perp^4$ Coulomb in the UV into a 
non-perturbative $1/q_\perp^3$ behaviour in the IR takes place, 
providing the bare minimum of 
``screening'' to make $\hat{q}$, the second moment of the kernel, IR finite. 
In the $q_\perp\sim gT$ range the non-perturbative curve differs appreciably 
from either the LO or NLO curves. 
\cite{Moore:2021jwe,Schlichting:2021idr,soudi} further analysed the impact 
of the non-perturbative kernel on MIEs, both for  infinite and finite media. 
They find that  an $\mathcal{O}(1)$ difference
between the MIE rates with NLO and non-perturbative kernels, 
a much larger effect than 
the improved opacity expansion compared to the full numerical
propagator.
See~\cite{kumar,ilia,Kumar:2020wvb,Grishmanovskii:2022tpb,soloveva,Soloveva:2021quj}
for other non-perturbative approaches to $\qhat$ and transport.

Let us now shift to \emph{quantum corrections}. In~\cite{Liou:2013qya,Blaizot:2013vha}
it was pointed out that radiative corrections to TMB from the recoil 
off the radiated gluon are responsible for a double-log enhanced, $\mathcal{O}(\alpha_s)$
correction to $\hat{q}$, arising from soft and collinear logarithms 
in the single-scattering regime, i.e.
\begin{equation}
    \delta\hat{q}=\frac{\alpha_s N_c}{\pi}\qhat_0 \int_{\tau_0}^\tau \frac{d\tau}{\tau} 
    \int_{\qhat_0\tau}^{k_\perp^2} \frac{dk_\perp^2}{k_\perp^2},
\end{equation}
where the $\qhat_0\tau$ boundary keeps the phase space 
away from the $k_\perp^2\sim \qhat \tau$ multiple scattering regime. 
These double-log correction can be understood as 
renormalizing $\qhat_0$, the LO value of $\qhat$. By 
promoting $\qhat_0$ to a variable and moving it (and the coupling) 
within the double integral, one obtains the resummation equation for these 
logs~\cite{Iancu:2014kga}. 
It was solved numerically and semi-analytically in~\cite{Caucal:2021lgf,Caucal:2022fhc,yacine},
finding how an initial $\qhat(\tau_0,k_{\perp\,0}^2)$ evolves to a 
$\qhat(\tau,k_{\perp}^2)$ resumming arbitrary numbers of long duration, quantum  
fluctuations. One of the main results is that the non-local nature 
of these fluctuations affects the UV tail of the TMB probability,
shifting it from its tree-level $\mathcal{P}\propto k_\perp^{-4}$ Coulomb 
form to a less steep $\mathcal{P}\propto k_\perp^{-4+2\alpha_s N_c/\pi}$. This 
larger probability for wider-angle scatterings corresponds to
more efficient diffusion. 

These quantum, radiative corrections are universal: as shown 
in~\cite{Blaizot:2014bha,Wu:2014nca,Iancu:2014kga}, they also arise 
in the case of a double MIE with overlapping formation times 
in the soft limit. Over the past years there has been an ongoing 
effort~\cite{Arnold:2015qya,Arnold:2016kek,Arnold:2016mth,
Arnold:2020uzm,Arnold:2021pin,Arnold:2022epx}
 to determine all these double-splitting real and virtual corrections,
beyond the soft limit. The underlying goal is understanding whether 
the assumed Markovian nature of the MIE kernel holds when used
to construct in-medium cascades. In the most recent developments
\cite{Arnold:2021mow,Arnold:2021pin,shahin} it was shown 
that, with some caveats, also the single-logarithmic radiative corrections 
determined in~\cite{Liou:2013qya} are universal, in that they apply also 
to double splitting, opening a promising pathway for their resummation.

\section{Kinetic theory, transport and thermalisation}
As mentioned, MIE and TMB are key ingredients of the kinetic description of QCD media.
They are complemented by drag, longitudinal momentum broadening and identity-change 
processes in the leading-order effective kinetic theory of QCD~\cite{Arnold:2002zm}.
We refer to \cite{Schlichting:2020lef,Dai:2020rlu,Ke:2020clc,ke,dai}
for recent applications of the kinetic framework to jet modifications. Here we 
instead connect the previous sections with
transport coefficients and thermalisation. 

Let us consider the shear viscosity $\eta$, which damps  flow gradients.
It is then clear that microscopic processes that isotropise momentum are 
dominant contributions. Hence, the direct isotropizing effect of TMB 
is more important for $\eta$ than its indirect effect 
as the driver of MIEs, as seen also in the LO determination 
of $\eta$~\cite{Arnold:2003zc}. These were recently extended to (almost)
NLO in~\cite{Ghiglieri:2018dib}. These corrections are large --- they reduce 
$\eta/s$ by a factor of a few in the phenomenological $T\sim \text{few}\;T_c$ range ---
and are dominated by the $\mathcal{O}(g)$ classical corrections to $\qhat$ of~\cite{CaronHuot:2008ni}
shown in Fig.~\ref{fig:kernel}. What should we make of this? Recently, 
it was pointed out in \cite{Muller:2021wri} that the problem might lie in the LO determination of TMB,
which then affects $\eta$ at LO. If that determination has eccessive screening --- see Fig.~\ref{fig:kernel}
--- it underestimates broadening, potentially explaining qualitatively this LO-NLO discrepancy.

\begin{figure}[t]
    \begin{center}
        \includegraphics[width=6cm]{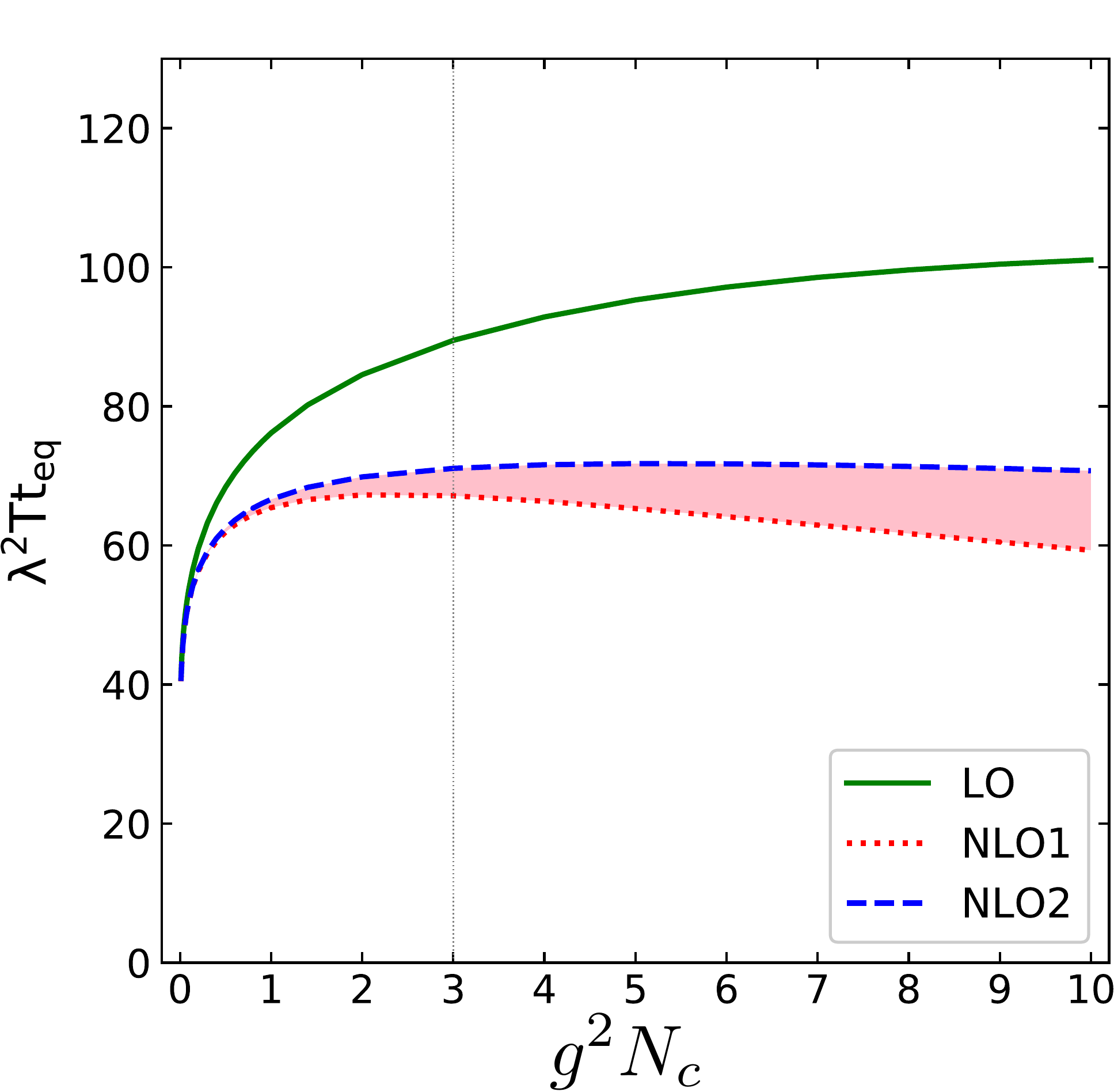}
    \end{center}
    \caption{Thermalisation time as a function 
    of the coupling $\lambda=g^2N_c$ 
    for an overouccupied
     initial condition. Figure adapted from \cite{Fu:2021jhl}.\label{fig:therm}}
\end{figure}
Several applications of kinetic theory to thermalisation have been discussed 
at this conference~\cite{Du:2020dvp,du,Brewer:2022vkq,Mikheev:2022fdl,bruno,plumari,Almaalol:2020rnu,almalool}.
Can similarly 
problematic perturbative expansions arise in these kinetic frameworks? What 
are the systematics of extrapolations to $\alpha_s=0.3$ ($g\approx 2$)?
In~\cite{Fu:2021jhl,fu}, NLO corrections to thermalisation for isotropic systems 
have been presented. Fig.~\ref{fig:therm} shows the LO and NLO thermalisation 
times for an overoccupied initial condition. Two different NLO collision operators have been 
constructed, which resum differently higher-order effects. Their spread, indicated 
by the shaded band, is a proxy for the size of even higher-order corrections. 
This band is smaller than its spread from the LO result, as expected for a convergent expansion. Moreover, the extrapolation 
to intermediate coupling seems controlled, with a 40\% correction for $g^2N_c=10$. However,
these isotropic sytems are by their nature insensitive to the isotropizing effect 
of TMB, which we argued to play a determining role in corrections to transport. 
It remains then to be understood how reliable the extrapolation could be in situations
typical of heavy-ion collisions with anisotropic initial conditions and expansions. 

In these systems,
plasma instabilities~\cite{Mrowczynski:1993qm,Romatschke:2003ms,Arnold:2003rq} ---
another classical phenomenon --- prevent at the moment consistent LO kinetic treatments.
Recently, the instability-subtracted TMB kernel, together with a recipe for dealing 
with the unstable modes, was provided in \cite{Hauksson:2021okc}, finding that 
anisotropy reduces the scattering kernel in the QGP phase. In the earlier 
glasma phase, large anisotropic TMB effects have been reported 
in~\cite{Ipp:2020nfu,Carrington:2022bnv,alina,schuh,meg}.

\textbf{Summary}: The reviewed  advances in the 
microscopic description of
 QCD media are instrumental in better quantification of theory uncertainties and  
in narrowing the gap between the QCD Lagrangian and phenomenology.

\bibliographystyle{myJHEP}
\bibliography{QM.bib}
\end{document}